# The semi-relativistic scattering states of the two-body spinless Salpeter equation with the Varshni potential model

O.J. Oluwadare[*] and K.J. Oyewumi[+]

[*]Department of Physics, Federal University Oye-Ekiti, Ekiti State, Nigeria.
[+]Theoretical Physics Section, Department of Physics, University of Ilorin, Ilorin, Nigeria.

**Abstract**
In this present work, the scattering state solutions of the Spinless Salpeter equation with the Varshni potential model were investigated. The approximate scattering phase shift, normalization constant, bound state energy, wave number and wave function in the asymptotic region were obtained. The behaviour of the phase shift with the two-body mass index $\eta$ were discussed and presented.

## 1. Introduction

The spinless Salpeter equation is a generalization of the Schrödinger equation in the relativistic regime [1-7], and it has been used to describe quark-antiquark interaction [3]. The two body effects have to do with relativistic considerations, special relativity, energy shifts of resonances, scattering states, bound states, and even the disappearance of excited bound state. These relativistic effects exist for a single Dirac mass in a potential model, described by Dirac Theory [8-10]. In the present study, we ignored the spin context and focus on the effects of two-body mass index $\eta$ on the scattering phase shift of the Varshni potential model.

The solution of this spinless Salpeter equation is limited to certain potentials due to the square of potential term in the equation. These potentials include: Yukawa potential [5, 6, 15, 16], generalized Hulthén potential [7, 12], Hulthén potential [11, 13], Cornell and Kratzer potential [17], Woods-Saxon potential [18-21], Coulomb potential [22] among the fewer ones. The solution of this equation with some potential models is of scientific interest as the semi-relativistic nature and two-body effects find their applications in particle and nuclear physics. The semi-relativistic phase shifts, wave numbers and wave functions in the asymptotic region of the scattering states can also be used in the theoretical prediction of many diatomic molecules.

*Corresponding author: oluwatimilehin.oluwadare@fuoye.edu.ng

[+] kjoyewumi66@unilorin.edu.ng

In this study, we chose Varshni potential for the fundamental role it plays in many field of physics including classical physics, modern physics, chemical and molecular Physics [23-26]. This potential model has been used to describe the bound states of the interaction systems. The Varshni potential was also studied by Lim (2009) using the 2-body Kaxiras-Pandey parameters. In his work, he reported that Kaxiras and Pandey used this potential to describe the 2-body energy portion of multi-body condensed matter [24].

In 2014, Arda and Sever [25] studied the pseudospin and spin symmetric solutions of Dirac equation with Hellmann potential, Wei-Hua potential and Varshni potential. The relativistic bound state energies and spinor wave function have been reported. The dependence of the energy eigenvalues on different quantum number pair $(n, \kappa)$ were presented. Oluwadare and Oyewumi (2017) also investigated the scattering states of DKP with Varshni potential and reported that the phase shift depends on total angular momentum $J$, Screening parameter $\beta$ and potential strengths $a$ and $b$ [26].

This work is presented as follows: Section 2 contains the scattering states of the two-body spinless Salpeter equation with the Varshni potential model. In Section 3, we presented discussion on the numerical and graphical results and the conclusion is given in Section 4.

## 2. Scattering states Solutions of the spinless Salpeter equation with Varshni potential

The spinless Salpeter equation for two-body particles interacting in a spherically symmetric potential in the center of mass system is given by [27, 28]:

$$\left[\sum_{i=1,2}\left(\sqrt{-\Delta + m_i^2} - m_i\right) + (V(r) - E_{n,l})\right]\chi(r) = 0, \quad \Delta = \nabla^2, \tag{1}$$

where $\chi(r) = R_{nl}(r)Y_{lm}(\theta, \varphi)$. For heavy interacting particles, we can write ($\hbar = c = 1$):

$$\psi_{nl}''(r) + \left[-l(l+1)r^{-2} + 2\mu(E_{n,l} - V(r)) + (\mu/\eta)^3(E_{n,l} - V(r))^2\right]\psi_{nl}(r) = 0, \tag{2}$$

where the transformation variable $R_{nl}(r) = \psi_{nl}(r)/r$ has been employed. The reduced mass $\mu = m_1 m_2/(m_1 + m_2)$ and mass index $= \mu[m_1 m_2/(m_1 m_2 - 3\mu^2)]^{1/3}$. $E_{n,l}$ is the semi-relativistic energy of the two particles with arbitrary masses $m_1$ and $m_2$. When the term with the mass index tends to zero, the solutions to Eq. (2) become non-relativistic.

The Varshni potential is written as [23-26]

$$V(r) = a\left[1 - \frac{b}{r}e^{-\beta r}\right], \tag{3}$$

where $r$ is the internuclear distance, $a$ and $b$ are the strengths of the potential and $\beta$ is the screening parameter which controls the shape of the potential energy curve.

To solve analytically for any $l \neq 0$ scattering states, we apply approximation scheme of the type [29-32 and the references therein]

$$\frac{1}{r^2} \approx \frac{\beta^2}{(1-e^{-\beta r})^2}. \tag{4}$$

The approximation has been reported to be valid for $\beta r \ll 1$. Inserting Eqs. (3) and (4) into Eq. (2) and applying a new variable $z = 1 - e^{-\beta r}$, one obtains

$$\psi_{nl}''(z) - \frac{1}{(1-z)}\psi_{nl}'(z) + \frac{1}{z^2(1-z)^2}[-w_1 z^2 + w_2 z - w_3]\psi_{nl}(z) = 0, \tag{5}$$

where

$$-w_1 = \frac{2ab\mu^3 E_{n,l}}{\eta^3 \beta} - \frac{2a^2 \mu^3 b}{\eta^3 \beta} + \frac{2\mu ab}{\beta} - \frac{a^2 b^2 \mu^3}{\eta^3} - l(l+1) - \frac{k^2}{\beta^2}, \tag{6}$$

$$w_2 = \frac{2\mu ab}{\beta} + \frac{2ab\mu^3 E_{n,l}}{\eta^3 \beta} - \frac{2a^2 \mu^3 b}{\eta^3 \beta} - \frac{2a^2 b^2 \mu^3}{\eta^3}, \tag{7}$$

$$-w_3 = -l(l+1) + \frac{a^2 b^2 \mu^3}{\eta^3}, \tag{8}$$

and $k = \sqrt{2\mu(E_{n,l} - a) + (\mu/\eta)^3 (E_{n,l} - a)^2 - l(l+1)\beta^2}$ is the asymptotic wave number.

Considering a trial wave function of the form:

$$\psi_{nl}(z) = z^\lambda (1-z)^{-i(k/\beta)} U_{nl}(z), \tag{9}$$

and substituting it into Eq. (5), enables us to arrive at the hypergeometric equation [33]

$$z(1-z)U_{nl}''(z) + \left[2\lambda - \left(2\lambda - 2i\frac{k}{\beta} + 1\right)z\right]U_{nl}'(z) + \left[\left(\lambda - i\frac{k}{\beta}\right)^2 + w_1\right]U_{nl}(z) = 0, \tag{10}$$

where we have used the following wave parameters:

$$\lambda = \frac{1}{2} + \sqrt{\frac{1}{4} + l(l+1) - \frac{a^2 b^2 \mu^3}{\eta^3}}, \tag{11}$$

$$\eta_1 = \lambda - i\frac{k}{\beta} - \sqrt{\frac{2ab\mu^3 E_{n,l}}{\eta^3 \beta} - \frac{2a^2 \mu^3 b}{\eta^3 \beta} + \frac{2\mu ab}{\beta} - \frac{a^2 b^2 \mu^3}{\eta^3} - l(l+1) - \frac{k^2}{\beta^2}}, \tag{12}$$

$$\eta_2 = \lambda - i\frac{k}{\beta} + \sqrt{\frac{2ab\mu^3 E_{n,l}}{\eta^3 \beta} - \frac{2a^2 \mu^3 b}{\eta^3 \beta} + \frac{2\mu ab}{\beta} - \frac{a^2 b^2 \mu^3}{\eta^3} - l(l+1) - \frac{k^2}{\beta^2}}, \tag{13}$$

$$\eta_3 = 2\lambda. \tag{14}$$

The radial wave function for the scattering states of the Varshni potential is obtained as:

$$\psi_{nl}(r) = N_{n,l}(1 - e^{-\beta r})^\lambda e^{ikr} {}_2F_1(\eta_1, \eta_2, \eta_3; 1 - e^{-\beta r}) \tag{15}$$

and $N_{n,l}$ is the normalization constant.

**The Semi-Relativistic Scattering Phase Shifts and Normalization Constant**

The semi-relativistic scattering phase shifts $\delta_l$ and normalization constant $N_{n,l}$ can be obtained by considering the well-known recurrence relation of hypergeometric function [33]

$$_2F_1(\eta_1,\eta_2,\eta_3;z) = \frac{\Gamma(\eta_3)\Gamma(\eta_3-\eta_1-\eta_2)}{\Gamma(\eta_3-\eta_1)\Gamma(\eta_3-\eta_2)} {_2F_1}(\eta_1;\eta_2;1+\eta_1+\eta_2-\eta_3;1-z)$$

$$+(1-z)^{\eta_3-\eta_1-\eta_2}\frac{\Gamma(\eta_3)\Gamma(\eta_1+\eta_2-\eta_3)}{\Gamma(\eta_1)\Gamma(\eta_2)} {_2F_1}(\eta_3-\eta_1;\eta_3-\eta_2;\eta_3-\eta_1-\eta_2+1;1-z). \quad (16)$$

With the aid of Eq. (16) and the condition that $_2F_1(\eta_1,\eta_2,\eta_3;0)=1$, when $r\to\infty$, leads to

$$_2F_1(\eta_1,\eta_2,\eta_3;1-e^{-\beta r}) \xrightarrow{r\to\infty} \Gamma(\eta_3)\left|\frac{\Gamma(\eta_3-\eta_1-\eta_2)}{\Gamma(\eta_3-\eta_1)\Gamma(\eta_3-\eta_2)} + e^{-2ikr}\left|\frac{\Gamma(\eta_3-\eta_1-\eta_2)}{\Gamma(\eta_3-\eta_1)\Gamma(\eta_3-\eta_2)}\right|^*\right|, \quad (17)$$

where we have used the following relations for simplicity:

$$\eta_3-\eta_1-\eta_2 = (\eta_1+\eta_2-\eta_3)^* = 2i(k/\beta), \quad (18)$$

$$\eta_3-\eta_2 = \lambda + i\frac{k}{\beta} - \sqrt{\frac{2ab\mu^3 E_{n,l}}{\eta^3\beta} - \frac{2a^2\mu^3 b}{\eta^3\beta} + \frac{2\mu ab}{\beta} - \frac{2a^2 b^2\mu^3}{\eta^3} - l(l+1) - \frac{k^2}{\beta^2}} = \eta_1^*, \quad (19)$$

$$\eta_3-\eta_1 = \lambda + i\frac{k}{\beta} + \sqrt{\frac{2ab\mu^3 E_{n,l}}{\eta^3\beta} - \frac{2a^2\mu^3 b}{\eta^3\beta} + \frac{2\mu ab}{\beta} - \frac{2a^2 b^2\mu^3}{\eta^3} - l(l+1) - \frac{k^2}{\beta^2}} = \eta_2^*. \quad (20)$$

Now, by taking

$$\frac{\Gamma(\eta_3-\eta_1-\eta_2)}{\Gamma(\eta_3-\eta_1)\Gamma(\eta_3-\eta_2)} = \left|\frac{\Gamma(\eta_3-\eta_1-\eta_2)}{\Gamma(\eta_3-\eta_1)\Gamma(\eta_3-\eta_2)}\right| e^{i\delta}, \quad (21)$$

and inserting this into Eq. (17), we have

$$_2F_1(\eta_1,\eta_2,\eta_3;1-e^{-\beta r}) \xrightarrow{r\to\infty} \Gamma(\eta_3)\left[\frac{\Gamma(\eta_3-\eta_1-\eta_2)}{\Gamma(\eta_3-\eta_1)\Gamma(\eta_3-\eta_2)}\right]$$

$$\times e^{-ikr}\left[e^{i(kr-\delta)} + e^{-i(kr-\delta)}\right]. \quad (22)$$

Thus, we obtain the asymptotic form of Eq. (15) for $r\to\infty$ as;

$$\psi_{nl}(r) \xrightarrow{r\to\infty} 2N_{n,l}\Gamma(\eta_3)\left[\frac{\Gamma(\eta_3-\eta_1-\eta_2)}{\Gamma(\eta_3-\eta_1)\Gamma(\eta_3-\eta_2)}\right]\sin\left(kr+\delta+\frac{\pi}{2}\right). \quad (23)$$

Using the boundary condition defined by Landau and Lifshitz [34], Eq. (23) reduces to

$$\psi_{nl}(\infty) \to 2\sin\left(kr+\delta_l-\frac{l\pi}{2}\right). \quad (24)$$

The phase shifts formula and the normalization constant are obtained, respectively as:

$$\delta_l = \frac{\pi}{2}(l+1) + arg\Gamma(2i(k/\beta)) - arg\Gamma(\eta_2^*) - arg\Gamma(\eta_1^*) \quad (25)$$

and

$$N_{n,l} = \frac{1}{\sqrt{\eta_3}}\left|\frac{\Gamma(\eta_1^*)\Gamma(\eta_2^*)}{\Gamma(2i(k/\beta))}\right|. \quad (26)$$

**The analytical properties of partial-wave S-matrix**

Here, we consider and discuss the following the property $\Gamma(\eta_3 - \eta_1)$ [34] as:

$$\eta_3 - \eta_1 = \lambda + i\frac{k}{\beta} + \sqrt{\frac{2ab\mu^3 E_{n,l}}{\eta^3 \beta} - \frac{2a^2\mu^3 b}{\eta^3 \beta} + \frac{2\mu ab}{\beta} - \frac{a^2 b^2 \mu^3}{\eta^3} - l(l+1) - \frac{k^2}{\beta^2}}. \tag{27}$$

The first order poles of $\Gamma\left(\lambda + i\frac{k}{\beta} + \sqrt{\frac{2ab\mu^3 E_{n,l}}{\eta^3 \beta} - \frac{2a^2\mu^3 b}{\eta^3 \beta} + \frac{2\mu ab}{\beta} - \frac{a^2 b^2 \mu^3}{\eta^3} - l(l+1) - \frac{k^2}{\beta^2}}\right)$ are situated at

$$\Gamma\left(\lambda + i\frac{k}{\beta} + \sqrt{\frac{2ab\mu^3 E_{n,l}}{\eta^3 \beta} - \frac{2a^2\mu^3 b}{\eta^3 \beta} + \frac{2\mu ab}{\beta} - \frac{a^2 b^2 \mu^3}{\eta^3} - l(l+1) - \frac{k^2}{\beta^2}}\right) + n = 0 \quad (n = 0, 1, 2, \dots). \tag{28}$$

By solving Eq. (28) using algebraic means, we obtained the bound state energy equation for the Varshni potential under the SSE equation as:

$$k^2 = -\beta^2 \left[\frac{(n+\lambda)^2 - \frac{2ab\mu^3 E_{n,l}}{\eta^3 \beta} + \frac{2a^2\mu^3 b}{\eta^3 \beta} - \frac{2\mu ab}{\beta} + \frac{a^2 b^2 \mu^3}{\eta^3} + l(l+1)}{2(n+\lambda)}\right]^2 = 0. \tag{29}$$

Table 1: Semi-relativistic scattering phase shift as a function of angular momentum quantum number $l$ for $a = b = 0.15, E = 1$.

| $l$ | $\delta_l$ for $\mu = 0.5, \eta = 0.79$ (Equal masses case) | $\delta_l$ for $\mu = 0.99, \eta = 0.99$ (Unequal masses case) |
|---|---|---|
| 0 | -3.20116 | -10.14667 |
| 1 | -0.48696 | -8.03766 |
| 2 | 3.94288 | -2.55423 |
| 3 | 1.44838 | 5.13545 |
| 4 | -1.48742 | 1.52129 |
| 5 | -4.77575 | -2.18885 |
| 6 | -8.36940 | -6.07785 |
| 7 | -12.23144 | -10.16280 |
| 8 | -16.33226 | -14.43993 |
| 9 | -20.64778 | -18.89947 |
| 10 | -25.15815 | -23.53029 |
| 11 | -29.84674 | -28.32149 |
| 12 | -34.69945 | -33.26291 |
| 13 | -39.70422 | -38.34528 |
| 14 | -44.85058 | -43.56023 |
| 15 | -50.12941 | -48.90018 |
| 16 | -55.53264 | -54.35831 |
| 17 | -61.05314 | -59.92842 |
| 18 | -66.68454 | -65.60491 |
| 19 | -72.42109 | -71.38266 |
| 20 | -78.25764 | -77.25701 |

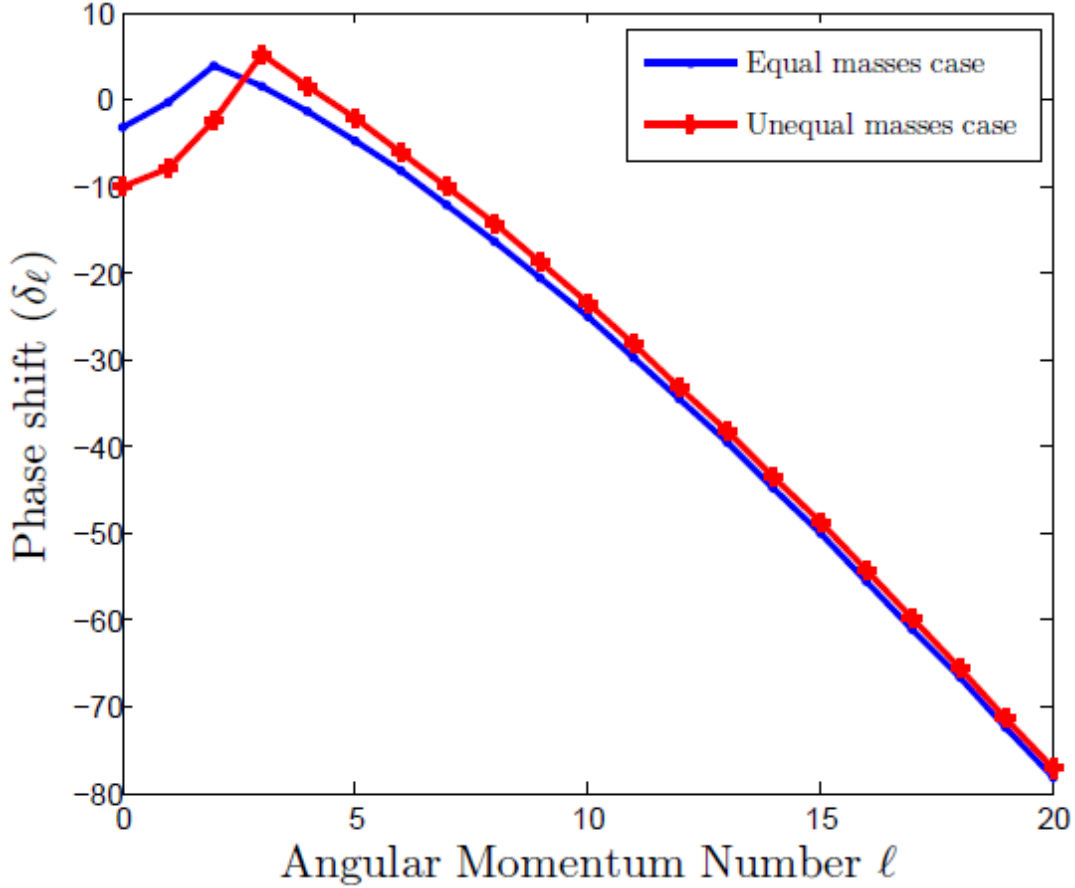

Figure 1: Semi-relativistic scattering phase shifts for the spinless Salpeter equation with the Varshni potential as a function of angular momentum quantum number $l$ with $a = b = 0.15$, $E_{n,l} = 1$.

The results are presented in the natural units $\hbar = c = 1$. The two-body effect in this work concerns with the different mass combinations for the same reduced mass $\mu$. For equal masses, we used $(\mu/\eta)^3 = 1/4$ and $\mu = m_1/2$ and for unequal masses case, we used $(\mu/\eta)^3 = 1$ and $\mu = m_1/100$. In all the cases, we consider $m_2 = E_{n,l} = 1$ and $m_1 = 1$ for the equal masses case only. The result in Table 1 and Figure 1 indicate an exponentially linear dependence of semi-relativistic scattering phase shift on angular momentum quantum number $l$. The essence of two-body effects is clearly seen in the Figure 1 where the behaviour of semi-relativistic scattering phase shift for unequal masses case seems better than that of equal masses case.

# 4    Conclusion

We have studied the semi-relativistic scattering states of two-body spinless Salpeter equation with Varshni potential. A suitable approximation scheme has been applied to overcome the effect of centrifugal term in the equation, and approximate scattering phase shift, normalization constant, the bound state energy equation at the pole of scattering amplitude, wave number and wave function in the asymptotic region have been obtained. It is seen from this work that two-body effects influence the behaviour of semi-relativistic scattering phase shift.

## Acknowledgment

The authors thanks Prof. K-E, Thylwe for making the program we used to analyze our results available.